\documentclass[prl,twocolumn,superscriptaddress,showpacs,floatfix]{revtex4-1}
\usepackage{epsfig}
\usepackage{color}
\usepackage{graphics}
\begin{document}

\title{The Inherent Structure Landscape Connection Between Liquids, Granular materials and the Jamming Phase Diagram.}

\author{S. S. Ashwin}
\affiliation{Department of Chemistry, University of Saskatchewan,
Saskatoon, Saskatchewan, S7N 5C9}
\affiliation{Department of Mechanical Engineering and Material Science \&
Department of Physics,\\
Yale University, New Haven, CT 06511}

\author{Mahdi Zaeifi Yamchi}
\affiliation{Department of Chemistry, University of Saskatchewan,
Saskatoon, Saskatchewan, S7N 5C9}

\author{Richard K. Bowles}
\affiliation{Department of Chemistry, University of Saskatchewan,
Saskatoon, Saskatchewan, S7N 5C9}
\email{richard.bowles@usask.ca}

\date{\today}

\begin{abstract}
We provide a comprehensive picture of the jamming phase diagram by connecting the athermal, granular ensemble of jammed states and the equilibrium fluid through the inherent structure paradigm for a system hard discs confined to a narrow channel. The J-line is shown to be divided into packings that are thermodynamically accessible from the equilibrium fluid and inaccessible packings. The J-point is found to occur at the transition between these two sets of packings and is located at the maximum the inherent structure distribution. A general thermodynamic argument suggests that the density of the states at the configurational entropy maximum represents a lower bound on the J-point density in hard sphere systems. Finally, we find that the granular and fluid systems only occupy the same set of inherent structures, under the same thermodynamic conditions, at two points, corresponding to zero and infinite pressures, where they sample the J-point states and the most dense packing respectively.
\end{abstract}

\maketitle
The way particles pack together influences the thermodynamic and mechanical properties of a wide variety of materials, including crystals, amorphous glasses and liquids, as well as many athermal materials such as granular sand piles and foams~\cite{con04}. Thermal and athermal materials have also been shown to have similar dynamic properties as the point at which the particles jam together into a mechanically stable packing is approached. The jamming phase diagram~\cite{ Liu98,ohern02}, which locates the jamming transition as a function of various thermodynamic and dynamic variables, was introduced in an attempt to describe the jamming phenomena of these systems within a unified conceptual framework. However, there is still considerable debate concerning the details of the jamming phase diagram~\cite{cia10a,hec10}, and in particular, how the J-point is connected to the thermodynamics and dynamics of these systems. For example, the J-point has been connected with the random close packing~\cite{song08}, while mean field studies~~\cite{kur07,kur09,cia10,zampar}, along with others~\cite{sas10}, suggest there a is a range of jamming densities, associated with a J-line, starting from the lowest density jammed packing related to the J-point and increasing up to the most dense, glass close packing state.

The goal of this letter is to explore the key elements of the jamming phase diagram through the analysis of a simple system for which both the statistical mechanics of the granular system, the equilibrium fluid and the connection between these two states of matter can be treated exactly within the inherent structure paradigm~\cite{gold69,deb01,stil64,spe98,spe01}. This allows us to move beyond the mean field results which are restricted to the study of jammed packing produced by quenching fluid states from above a density, $\phi_d$, where the fluid phases splits into an exponential number of states and particles are unable to diffuse out of their respective cages. Our results show that the J-line extends over a broader range of densities than expected. Furthermore, the J-line is divided into packings that are thermodynamically accessible or inaccessible from the equilibrium fluid. We also find that the J-point occurs at the maximum of the configurational entropy and marks the transition between the accessible and inaccessible packings.

\begin{figure}[t]
\includegraphics[width=3.2in]{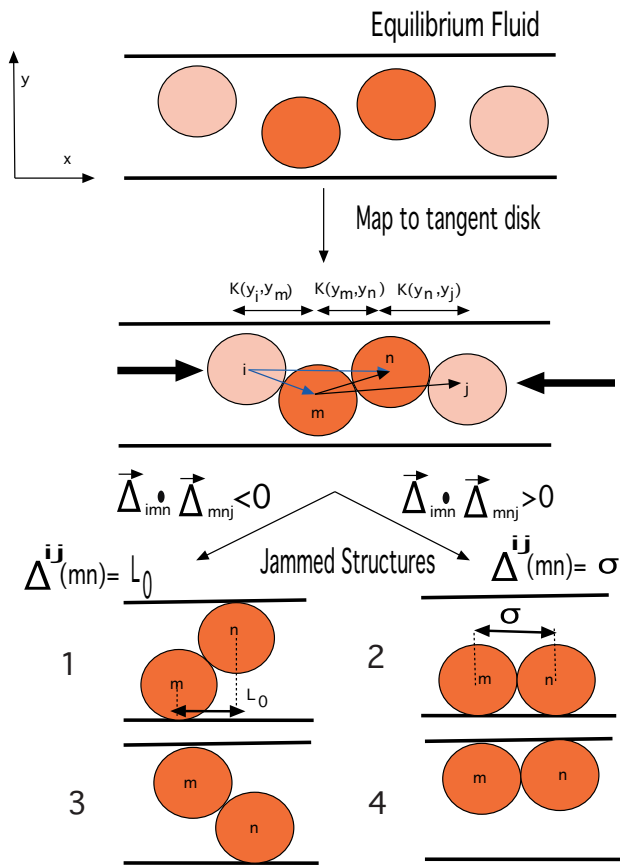}
\caption{Analytical quench connecting equilibrium fluid configurations of discs $i,m,n,j$ to the most dense (1,3) and least dense (2,4) local packing arrangements. Equilibrium configuration of 4 discs are initially mapped to a tangent discs configuration by compression in the x-axis. The local arrangement of the discs contained in the product of the kernel $K(y_i,y_j)$ and the dot product of vectors (Eq~\ref{eq:tri}) maps the central discs to their jammed configuration. See text for more details.}
\label{fig:map}
\end{figure}

We begin by generating the full ensemble of jammed states for a system of $N$, two dimensional ($2D$) hard discs of diameter $\sigma$, confined to a narrow channel by two hard walls separated by a distance $H/\sigma<1+\sqrt{3/4}$. In $D$-dimensions, a spherical particle is locally jammed if it has at least $D+1$ rigid contacts arranged such that they are not all within the same hemisphere.  However, the local jamming of all the particles in a structure is a necessary, but not sufficient, condition to ensure collective jamming because the concerted motion of a group of particles may allow the structure to collapse~\cite{tor01,don05}.  The confinement of the present model prevents the discs from passing each other, which eliminates the possibility of collective motions of particles unjamming the packings and allows us to count all the collectively jammed structures by simply considering local packing constraints. It also ensures that each disc can only interact with its nearest neighbors and the wall. As a result, there are only four particle configurations that satisfy the local jamming constraints, two dense configurations (denoted 1 and 3) and two  open, defect type configurations, denoted 2 and 4 (see Fig.~\ref{fig:map}). The volume associated with each configuration is given by $Hl_{ij}$, where the $l_{ij}$ is the longitudinal distance between neighboring discs' centres, with $l_{i,1}=l_{i,3}= [H(2\sigma-H)]^{1/2}$ and $l_{i,2}=l_{i,4}=\sigma$. Any jammed configuration can now be identified by a list of neighboring bonds.  However, it is important to note that configurations containing neighboring defects are not allowed as the central disc in the local arrangement is unjammed. Furthermore, all the jammed states of the model are isostatic~\cite{rkb10}.

Following the statistical mechanics approach to granular materials suggested by Edwards~\cite{ede,rkb10}, the quasi-one-dimensional nature of the model allows us to use the transfer matrix approach to construct the ensemble of jammed states, taking bond incompatibilities into account~\cite{ash09}. For a fixed $N$, the volume of the system will fluctuate depending on the number of type 2 and 4 states in the configuration so we introduce a longitudinal pressure $P_{L}$, as a conjugate to the volume, and fix the system at a constant temperature, $T$. The transfer matrix then takes the form:
\begin{equation}
M=\left(
\begin{array}{llll}
 0& M_{1,2} & M_{1,3} &0 \\
 0 & 0 & M_{2,3} & 0  \\
 M_{3,1} & 0& 0 & M_{3,4}  \\
 M_{4,1}& 0& M_{4,3} & 0 \\
\end{array}
\right)
\label{eq:tm1}
\end{equation}
where $M_{ij}=C_{ij}\exp(-\beta P_{L}h_0 l_{ij})$, where $h_0=H-\sigma$. The exponential term is the Gibbs measure appropriate for the $N,P_{L},T$ ensemble and $C_{ij}$ is zero when the two bonds are incompatible and one otherwise. $\beta=1/kT$, where $k$ is the equivalent to the Boltzmann constant in the jammed ensemble. In the thermodynamic limit, the partition function for the system is given by $\Delta(N,P_{L},T)=NkT\ln(\lambda)$, where $\lambda$ is the largest eigenvalue of $M$.  The configurational entropy, which is directly related to the number of jammed packings, $N_J$, by $S_c=k\ln N_J(\phi_J)$, and $\phi_J$ are given 
\begin{equation}
S_c/Nk=\ln\lambda+ T\partial (\ln \lambda)/\partial T\mbox{ .}\\
\label{eq:s1}
\end{equation}
\begin{equation}
\phi_J=\frac{N\pi\sigma^2}{4HL_J}=-\frac{\pi\sigma^2}{4kT\partial(\ln \lambda)/\partial P_{L}}\mbox{ ,}\\
\label{eq:phij1}
\end{equation}

\begin{figure}[t]
\includegraphics[width=3.5in]{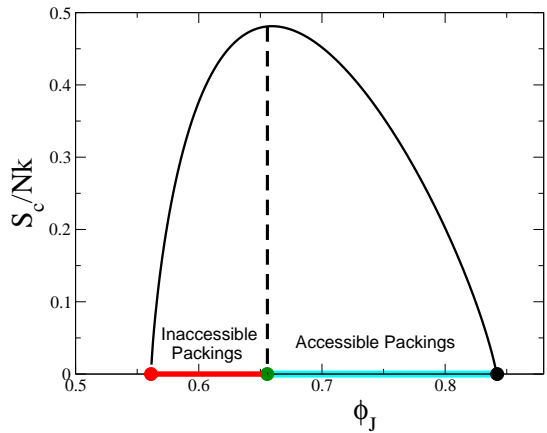}
\caption{$S_c/Nk$ versus $\phi_J$ for $H/\sigma=1.866$ showing the J-line. The cyan line highlights the set of $\phi_J$ belonging to the accessible jammed states which is bracketed by the most dense (black point) and the J-point (green point), which also coincides with the maximum in $S_c$. The red line indicates the inaccessible packings leading to the least dense packing (red point).}  
\label{fig:djs}
\end{figure}

\begin{figure}[t]
\includegraphics[width=3.5in]{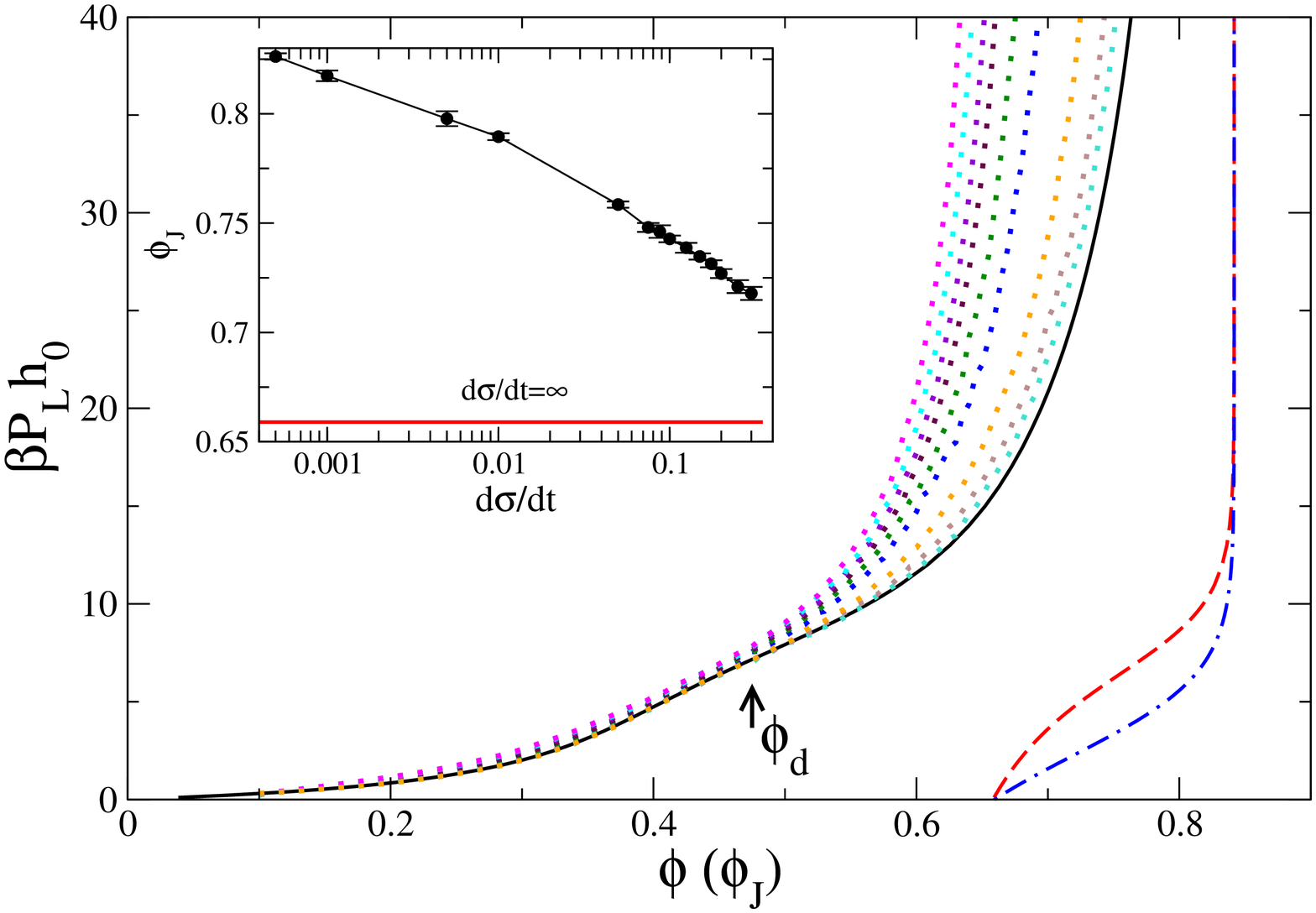}
\caption{Equations of state  for $H/\sigma=1.866$. EOS for granular ensemble plotting $\beta P_{L}h_0$ as a function of $\phi_J$ (blue dot-dash line). EOS for equilibrium thermal fluid plotting $\beta P_{L}h_0$ as a function of $\phi$ (black solid line). $\phi_J$ of basins sampled by equilibrium fluid (red dashed line). EOS for fluid compressed using the LS method with different $\partial\sigma/\partial t$ (dotted lines). $\phi_d\approx 0.48$ obtained from MD simulations. Insert: $\phi_J$ as a function of  $\partial\sigma/\partial t$.}
\label{fig:eos_liq}
\end{figure}

Fig.~\ref{fig:djs} shows $S_c/Nk$ as a function of $\phi_J$ for the case when $H/\sigma=1.866$, obtained from a parametric plot of Eqs. \ref{eq:phij1} and \ref{eq:s1} with respect to $P_{L}$. The same distribution of states for this system can be obtained using a combinatorial approach~\cite{rkb06} and gives $S_c/Nk=(1-\theta)\ln(1-\theta)-\theta\ln\theta-(1-2\theta)\ln(1-2\theta)$, where $\theta$ is the mole fraction of defects (type 2 and 4 bonds). The advantage of the present method is that it allows us to follow how the granular system explores the packing landscape as a function of the externally applied pressure and we have plotted the full equation of state (EOS) in Fig~\ref{fig:eos_liq}. While there is no internal pressure from the particles in a granular system, it is still necessary for it to do work against $P_L$ if the system expands to samples less dense states, but there are fewer high density basins, so the balance between these two competing elements results in the ``equilibrium" condition for the granular system. In the limit that $P_{L}h_0\rightarrow\infty$, $S_c/Nk\rightarrow 0$ as the system moves towards the most dense state with $\phi_J=0.842$, while as $P_{L}h_0\rightarrow0$, the system samples the jammed states associated with the maximum in $S_c$ where $\theta=1/2-\sqrt{5}/10$ and $\phi_J=0.659$. If we allow the pressure to go below zero, we can sample less dense packings and the system enters the least dense jammed state with $\phi_J=0.561$, $S_c/Nk=0$, as $P_{L}h_0\rightarrow-\infty$. However, the hard particles have no attractive component to their interaction potential that could sustain a negative pressure, suggesting the packings below the $S_c$ maximum are thermodynamically inaccessible.

We now examine how the equilibrium fluid for this system samples the landscape. The exact partition function for this model was originally solved by Barker\cite{bar62}, but we make use of transfer matrix solution by Kofke {\it et al}~\cite{kof93}. If the positions of the discs are fixed in the $y$-direction, the configurational integral in the $x$-direction can be simply treated as a Tonks gas\cite{tonks36}. Using a Laplace transform,
 the volume dependence is removed, leaving the transfer integral for the partition function $Z$ in the NPT ensemble:
\begin{equation}
Z = \frac{1}{\Lambda^{DN}(\beta P_L)^{N+1}} \int dy K^{N}(y,y)\mbox{. }\\
\label{eq:z1}
\end{equation}
Here $\Lambda$ is the thermal wavelength, $P_L$ is the longitudinal pressure and $K(y_1,y_2)=\exp[-P_L h_0 L_x(y_1,y_2)]$, with $y_1$ and $y_2$ being the y-coordinates of two adjacent discs in contact. $L_x$ is the  projection of the distance between the two contacting discs along the $x$-axis and is a function of $y_1,y_2$. Solving the eigenvalue problem associated with Eq. \ref{eq:z1}~\cite{kof93} yields the equilibrium equation of state for the fluid plotted in Fig. \ref{fig:eos_liq}.

To quench the equilibrium fluid to its inherent structure, we take advantage of the information contained within the matrix $K$ regarding the geometry of adjacent tangent discs. Starting from an equilibrium configuration, we translate the discs along the $x$-axis only so that discs  are in contact with their nearest neighbors. Fig.~\ref{fig:map} shows that the type of bond (1,3 or 2,4) between the two central discs (mn) that will result from further compression can be determined from the sign of the product of areas made from the triangles created by particles $i,m,n$ and $m,n,j$. The geometry of the four discs is contained in the chain product matrix $ K(y_i,y_m)K(y_m,y_n)K(y_n,y_j)$. The product {\it area-vector-product} rule, for  triangles $\vec{\triangle}_{imn}$ and $\vec{\triangle}_{mnj}$, that determine the nature of the bond is,
\begin{eqnarray}
\vec{\triangle}_{imn}\bullet\vec{\triangle}_{mnj}& > 0 & \text{ bond  mn} \hspace{1mm} \Delta_{x}^{ij}(mn) =l_{k,1}\nonumber\\
\vec{\triangle}_{imn}\bullet\vec{\triangle}_{mnj}& < 0&  \text{ bond  mn} \hspace{1mm} \Delta_{x}^{ij}(mn) =\sigma\mbox{.}
\label{eq:tri}
\end{eqnarray}

We can now define a new transfer matrix $G$, whose elements are weighted by the bonds they would jam to under the jamming criterion,
\begin{equation}
G(i_2,i_3)=\sum_{i=i_1,i_4}K(i_1,i_2)K(i_2,i_3)K(i_3,i_4)\exp[\gamma\Delta^{ij}_x(mn)]\mbox{.}\\
\label{eq:g23}
\end{equation}
For a system with periodic boundary conditions and $N-2$ particles, the volume is given by,
\begin{equation}
V^{inh}_{N-2} =\lim \limits_{\gamma\rightarrow 0} \partial Log [Tr (G)]/ \partial \gamma\mbox{.}\\
\end{equation}

This method represents an ideal, infinitely-fast, non-equilibrium quench of the liquid to its inherent structure and amounts to a Stillinger-map, where every configuration is mapped uniquely to the jammed state in the basin. Consequently, each jammed state is weighted by the volume of configuration space  in the basin. Fig.~\ref{fig:eos_liq} shows both the equilibrium EOS for the fluid as a function of $\phi$ and the $\phi_J$ of the inherent structure sampled by the fluid. We can now compare how the equilibrium fluid and jammed system sample the same inherent structure landscape as a function of the externally applied pressure. While they exhibit the same general trend by moving to deeper inherent structures as $P_L$ increases, the two systems visit different basins at the same pressure. The equilibrium properties of the fluid arise from competition between the configurational entropy (the number of basins) and the free volume of a single basin, while the jammed ensemble has no free volume contribution. As a result, the fluid samples lower density basins than the jammed system at the same pressure.

However, the fluid and the jammed ensemble do sample the same set of basins at two particular values of $P_L$. In the limit that $P_L\rightarrow\infty$, they both move towards the most dense state. There is only a single most dense structure, so the configurational entropy goes to zero, but there is no ideal glass transition in the fluid as this only occurs at $P_L\rightarrow\infty$. The two systems also sample the basins associated with the maximum in $S_c$ as $P_L\rightarrow 0$, i.e. as the fluid approaches the ideal gas state. 

A simple thermodynamic argument can help us understand which jammed states are thermodynamically accessible in hard sphere systems in general. The entropy of a hard particle system at $\phi$, can be written as $S_f(\phi)/Nk=\ln N_g(\phi_J)+\ln Q_g(\phi,\phi_J)$, where $Q_g(\phi,\phi_J)$ is the configuration space of a single inherent structure basin that maps to $\phi_J$. At equilibrium, the system samples the set of basins with $\phi_J$ that satisfy, 
\begin{equation}
(\partial S_f/\partial \phi_J)_{\phi}=0\mbox{ .}\\
\label{eq:scid}
\end{equation} 
The free volume equation of state~\cite{don05b} and the equation of state for constrained glasses~\cite{spe98,spe01} suggest that $(\partial Q_g(\phi,\phi_J)/\partial \phi_J)_{\phi}\ge0$. On the other hand, we generally expect there to be a single maximum in $N_g(\phi_J)$, with only a few very high or very low density states, because of the coupling between density and structural order~\cite{tor00} and estimates of hard sphere packing distributions generally yield Gaussian type distributions~\cite{spe98,spe01}.
Combined, these conditions necessarily imply the equilibrium system is unable to sample basins with a $\phi_J$ lower than those at the $S_c$ maximum. If the system did sample states with $\phi_J$ lower than the maximum, we would have $(\partial N_g(\phi_J)/\partial \phi_J)>0$ and the equilibrium condition could never be satisfied. If the ideal gas does sample the associated with the $S_c$ maximum, then $(\partial Q_g(\phi,\phi_J)/\partial \phi_J)_{\phi}=0$, which suggests that the basins, for all the glasses become the same in the low density limit. This is true for the quasi-one-dimensional system studied here, and for the one-dimensional system of non-additive hard rods, where at $\phi=0$ every particle in the every glass becomes a point caged by its two neighboring points on the line.  Speedy~\cite{spe99} also found, for a binary mixture of hard discs in two dimensions, that the difference in entropy between the free ideal gas and the ideal gas constrained to a single basin is independent of the basin's $\phi_J$. However, a recent study~\cite{ash11} of disc packings involving a small number of particles suggest the the basin volumes of individual packings have complex geometries at low densities suggesting it is not obvious that all packings will have the same basin volume in the ideal gas state. Nevertheless, Eq~\ref{eq:scid}, in the limit $\phi\rightarrow 0$, defines the lowest density jammed packing accessible to the equilibrium fluid, which also provides a thermodynamic definition of the J-point.

We now compare our ideal inherent structure mapping with a more traditional compression scheme applied to hard sphere particles. We performed molecular dynamics simulations of our system using $N=10000$ discs. Starting from a random, low density configuration with $\phi=0.05$, the discs were compressed using a modified version of the Lubachevsky and Stillinger~\cite{ls90} scheme that ensures $H/\sigma$ remains constant as the discs are expanded at a rate of $d\sigma/dt$. The EOS for the system under different compression rates are plotted in Fig~\ref{fig:eos_liq}. The compressions follow the equilibrium EOS at low densities because the compression scheme allows the system to move between basins as it evolves. Eventually the fluid falls out of equilibrium at higher densities when caging effects start to become important and the system becomes trapped in a glassy state consisting of a single basin on the inherent structure landscape. Continued compression leads to a jammed state where the pressure diverges. The jamming density of the glasses as a function of compression rate (see insert Fig~\ref{fig:eos_liq}) was obtained by counting the number of defects in the glass and taking averages over 20 independent runs at each $d\sigma/dt$. Not surprisingly, the jamming density increases with decreasing $d\sigma/dt$ as slower compression rates allow the system to remain in equilibrium longer. To reach the most dense state, the system would have to be compressed sufficiently slowly that it was in equilibrium at all points along the trajectory. Finally we note that the density at which the system starts to fall out of equilibrium as a result of compression in Fig.~\ref{fig:eos_liq} represents the $\phi_d$ used in the mean field approaches.  The $\phi_J$s of packings generated from fluid configurations at this density are approximately 0.72, but our exact results show that there are many more jammed configurations with a lower $\phi_J$.

To conclude, our results show that the behavior of fluids, glasses and jamming can all be related through the underlying inherent structure landscape. Our model gives rise to a distribution of packings that is essentially the same as that speculated to exist by Ciamarra et al~\cite{cia10}. However, we also find that the fluid and jammed ensembles are connected thermodynamically on the jamming phase diagram at two points corresponding the least and most dense states accessible to the system from the equilibrium fluid. The least dense states accessible to the system occur at $P_L=0$ and are associated with the maximum in $S_c$. These states are connected to ideal gas states by an ideal, infinitely fast compression and mark the transition between the thermodynamically accessible and inaccessible packings. The most dense state occurs at $P_L=\infty$, with $S_c=0$, and is only accessible from an ideal gas starting configuration by an infinitely slow compression.

\end{document}